\documentclass[preprint,12pt]{elsarticle}

\usepackage{amssymb}
\usepackage[cmex10]{amsmath}
\usepackage{array}
\usepackage{mdwmath}
\usepackage{mdwtab}
\usepackage{eqparbox}
\usepackage{url}
\usepackage{cite}
\usepackage{amsmath,amssymb,amsfonts}
\usepackage{algorithm}
\usepackage{algpseudocode}
\usepackage{textcomp}
\usepackage{xcolor}
\usepackage{amsthm}

\journal{}

\begin{document}

\begin{frontmatter}

%% Title, authors and addresses

%% use the tnoteref command within \title for footnotes;
%% use the tnotetext command for theassociated footnote;
%% use the fnref command within \author or \affiliation for footnotes;
%% use the fntext command for theassociated footnote;
%% use the corref command within \author for corresponding author footnotes;
%% use the cortext command for theassociated footnote;
%% use the ead command for the email address,
%% and the form \ead[url] for the home page:
%% \title{Title\tnoteref{label1}}
%% \tnotetext[label1]{}
%% \author{Name\corref{cor1}\fnref{label2}}
%% \ead{email address}
%% \ead[url]{home page}
%% \fntext[label2]{}
%% \cortext[cor1]{}
%% \affiliation{organization={},
%%             addressline={},
%%             city={},
%%             postcode={},
%%             state={},
%%             country={}}
%% \fntext[label3]{}

\title{Auto-optimization of Energy Generation for Wave Energy Converters with Active Learning}

%% use optional labels to link authors explicitly to addresses:
%% \author[label1,label2]{}
%% \affiliation[label1]{organization={},
%%             addressline={},
%%             city={},
%%             postcode={},
%%             state={},
%%             country={}}
%%
%% \affiliation[label2]{organization={},
%%             addressline={},
%%             city={},
%%             postcode={},
%%             state={},
%%             country={}}

\author[label1]{Siyang Tang} %% Author name
\author[label1]{Wen-Hua Chen}
\author[label1]{Cunjia Liu}

%% Author affiliation
\affiliation[label1]{organization={Department of Aeronautical and Automotive Engineering, Loughborough University},%Department and Organization
            addressline={Epinal Way}, 
            city={Loughborough},
            postcode={LE11 3TU}, 
            state={Leicestershire},
            country={U.K.}}

%\affiliation[label2]{organization={Research Centre for Low Altitude Economy, The Hong Kong Polytechnic University},%Department and Organization
%            addressline={11 Yuk Choi Road, Hung Hom, Kowloon}, 
%            city={Hong Kong},
%            postcode={EF701}, 
%            state={Hong Kong},
%            country={China}}

%% Abstract
\begin{abstract}
%% Text of abstract
This paper presents an auto-optimization control framework for wave energy converters (WECs) to maximize energy generation under unknown and changing ocean conditions. The proposed control framework consists of two levels. The high-level controller operating at a longer time scale aims to maximize the average energy generation over several wave periods. The generated Power Take-Off (PTO) profile as the reference for the low-level physical system to follow. The new auto-optimization process leverages the parameterization of the non-stationary operation condition in WECs, establishing the relationship between the average energy generation and the key design parameters of the PTO force subject to the unknown wave parameters. The high-level controller is designed based on the concept of Dual Control for Exploration and Exploitation (DCEE) to quickly learn the unknown wave parameters by actively probing the ocean condition, while generating the optimal PTO profile. During this process, the uncertainty of the estimated wave condition is quantified and embedded in the optimization cost function to enable active learning. Simulation results under unknown regular and irregular waves demonstrate the effectiveness and robustness of this novel auto-optimization WEC systems with active learning, outperforming model predictive control, extremum seeking and classic Bang-Bang control approaches. 
\end{abstract}

%%Graphical abstract
%\begin{graphicalabstract}
%\includegraphics{grabs}
%\end{graphicalabstract}

%% Keywords
\begin{keyword}
%% keywords here, in the form: keyword \sep keyword
dual control \sep  active learning \sep wave energy converter \sep auto-optimization control

%% PACS codes here, in the form: \PACS code \sep code

%% MSC codes here, in the form: \MSC code \sep code
%% or \MSC[2008] code \sep code (2000 is the default)

\end{keyword}

\end{frontmatter}

%% Add \usepackage{lineno} before \begin{document} and uncomment 
%% following line to enable line numbers
%% \linenumbers

%% main text
%%

%% Use \section commands to start a section
\section{Introduction}

Renewable alternatives to fossil fuels are gaining momentum in response to the growing urgency of climate change. With approximately 71\% of the Earth's surface covered by oceans, these bodies of water offer an immense and largely untapped source of clean energy. Among various marine energy sources, wave energy stands out due to its superior power density compared to solar and wind energy~\citep{cruz2007ocean}. Wave Energy Converters (WECs), which transform ocean wave energy into electricity, play a central role in harnessing this resource.

Automatic control is pivotal for optimizing energy extraction in WECs by manipulating the Power Take-Off (PTO) mechanism. Numerous control strategies have been developed to maximize energy capture~\citep{nolan2005optimal, babarit2006optimal, anderlini2018reactive, garcia2017real, parrinello2020adaptive}. Among them, complex conjugate control determines the optimal velocity trajectory based on the frequency characteristics of the hydrodynamic excitation force. Specific implementations include derivative latching~\citep{nolan2005optimal} and de-latching control~\citep{babarit2006optimal}. Reactive control further extends passive strategies by adjusting both PTO resistance and reactance~\citep{garcia2017real}. More recently, data-driven approaches such as reinforcement learning~\citep{anderlini2018reactive} and neural networks~\citep{anderlini2017reactive} have been explored to adapt to unknown and time-varying sea conditions.

Model Predictive Control (MPC) and its variants~\citep{richter2012nonlinear, zhan2018adaptive} are recognized for their ability to handle system constraints while achieving promising control performance. Substantial efforts have been made to tailor MPC methods for WECs, as surveyed in~\citep{faedo2017optimal}. With the development of wave forecasting technologies~\citep{fusco2011study}, predictive control methods can incorporate future wave information to enhance energy harvesting~\citep{li2014model}. However, the high computational burden of MPC often limits its applicability in real-time scenarios~\citep{faedo2017optimal}. Additionally, the control performance index for WECs is frequently non-convex, posing further challenges to optimization~\citep{ringwood2014control}.

Wave energy generation is fundamentally different from solar and wind power systems. Whereas the latter typically operate around a static maximum power point~\citep{faedo2021energy}, WECs must synchronize their motion with the oscillatory nature of ocean waves. As such, their optimal operating condition is inherently non-stationary and periodic, and cannot be described by a fixed point in the control space~\citep{ringwood2014energy}. This unique challenge motivates the need for autonomous control systems capable of learning and adapting to changing ocean conditions.

In parallel to these physically motivated strategies, learning-based approaches have gained increasing attention. A notable example is the recent work on deep reinforcement learning (DRL)-based non-causal control for wave energy conversion~\citep{wang2024deep}, which proposes a data-driven scheme that combines wave prediction with adaptive control gain tuning. Such approaches showcase the potential of learning-based intelligence in WEC control.
This paper investigates the problem of maximizing wave energy generation from the perspective of autonomous learning and adaptation. We consider the WEC as an intelligent agent operating in a partially unknown and dynamically changing environment, with the goal of maximizing its energy output by actively learning the characteristics of the wave field. Ideally, such a control system should not only adapt to the environment but also explore it efficiently to accelerate the learning process. This requires the WEC to interact with ocean waves in a way that both improves energy capture and yields informative data about the wave conditions.

To achieve this, we adopt the Dual Control for Exploitation and Exploration (DCEE) framework~\citep{chen2021dual, chen2022perspective}, a recently developed approach in autonomous control. Originally introduced for source-seeking problems in atmospheric dispersion~\citep{hutchinson2019unmanned}, DCEE enables an agent to simultaneously exploit current knowledge to maximize performance while exploring the environment to reduce uncertainty through active learning. The framework has demonstrated promising results in other energy optimization tasks under environmental uncertainty, such as maximizing photovoltaic (PV) power generation~\citep{li2023dual}, and has been shown to outperform conventional reinforcement learning in some scenarios~\citep{li2000autonomous}.

Inspired by these successes, this paper applies DCEE to the WEC system to tackle the challenge of maximizing energy generation under uncertain and time-varying sea conditions. 
However, a key distinction lies in the non-stationary nature of the optimal operating condition in WECs—something not encountered in prior DCEE applications.

To address this challenge, we propose to parameterize the periodic behavior of the optimal PTO profile using trigonometric functions, consistent with the wave-induced motion of the buoy. While previous works have used truncated Fourier series or orthogonal polynomial projections to approximate system states and controls~\citep{bacelli2011control, faedo2018finite}, we adopt a more concise parametrization based on key trigonometric parameters. Through analytical derivation, we establish a functional relationship between the average generated energy and these key parameters, thereby transforming the problem of control in an unknown environment into one of learning the wave characteristics from observed energy output.

Unlike prior approaches that estimate wave parameters such as frequency using spectral or short-horizon estimation methods~\citep{fusco2010short, davis2020wave}, our framework employs Bayesian inference to estimate these parameters while simultaneously quantifying the associated uncertainty in real time. This uncertainty estimation is crucial to the DCEE strategy, where control actions are optimized not only to track the estimated optimal behavior (exploitation), but also to probe the wave conditions more effectively (exploration). By leveraging the DCEE framework and the proposed parametrization, we develop a novel auto-optimization control system capable of synchronizing WEC dynamics with unknown and varying wave fields to maximize energy capture.

The contributions of this paper are threefold. 
\begin{enumerate} 
\item We introduce a novel two-level autonomous control system architecture for WECs, specifically designed to address the challenges posed by the dynamic and immeasurable ocean environment. In this architecture, the high-level aims to actively learn the ocean condition and then generate the PTO profile accordingly. It consists of a particle filter for key parameter estimation and an intelligent search controller designed using DCEE, operating at a large time scale (e.g., 10 wave periods or ever more).  The low-level controller operating at much faster speed drives the WEC to follow the profile set by the high-level controller by altering PTO force in the physical system.
\item This paper transforms maximizing energy generation in an unknown ocean environment into a problem of searching unknown optimal operation condition. To speed up the learning process of the unknown ocean condition, we introduce the recently developed DCEE framework where, by carefully interacting with the wave, the WEC can generate more informative data/measurements that speed up the learning process. This active learning approach can achieve the optimal trade-off between exploitation and exploration. 
\item This work extends the DCEE framework to systems with non-stationary optimal conditions, a novel contribution. By parametrizing the periodic motion of the buoy and establishing its relationship with both PTO control and wave forces, we enable DCEE to operate effectively in this new class of control problems. 
\end{enumerate}

The remainder of this paper is organized as follows: Section~\ref{Problem formulation} formulates the WEC control problem and the parametrization approach. Section~\ref{DCEE} presents the DCEE-based algorithm and its implementation via particle filters and search methods. Section~\ref{simulation} provides simulation results comparing the proposed approach with benchmark control methods including extremum seeking, MPC, and Bang-Bang control. Finally, Section~\ref{conclusion} concludes the paper.

\section{Problem formulation}
\label{Problem formulation}
\subsection{Parametrization of ideal PTO profile}
A schematic diagram of the point absorber-type WEC is shown in Figure~\ref{point absorber}. It typically consists of a floating buoy that oscillates vertically with incident ocean waves, relative to a fixed or semi-fixed reference such as a mooring base or seabed anchor. The relative motion between the buoy and the fixed structure drives a Power Take-Off (PTO) system—commonly realized through hydraulic or electromechanical components—to convert mechanical wave energy into electrical power~\citep{cruz2007ocean}. The instantaneous power extracted is given by~\citep{yu1995state}:

\begin{equation}
    P(t) = -\mathcal{F}_{u}(t) v(t),
    \label{power}
\end{equation}

where $v(t)$ is the velocity of the buoy’s midpoint and $\mathcal{F}_{u}(t)$ is the control force applied by the PTO system.

The hydrodynamic model of the WEC can be expressed as~\citep{li2014model}:

\begin{figure}[ht]
\centerline{\includegraphics[width=0.4\textwidth]{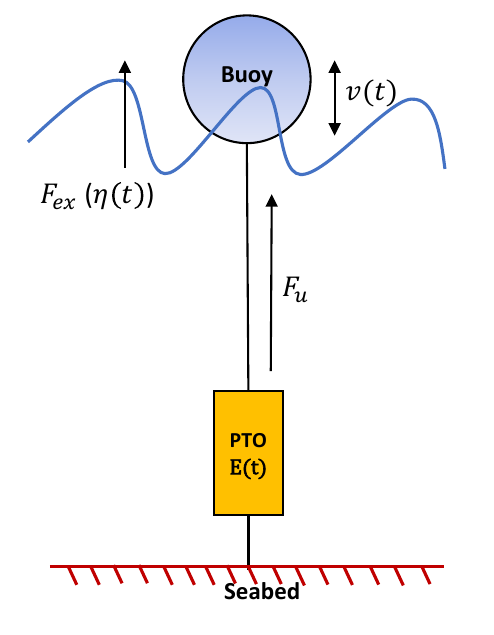}}
\caption{Schematic diagram of the point absorber.}
\label{point absorber}
\end{figure}

\begin{equation}
   m\dot{v}(t) + h_{r} v(t) + K x(t) = h_{ex} \eta(t) + \mathcal{F}_{u}(t),
   \label{wec motion simple}
\end{equation}
where $m$ is the sum of the buoy mass and added mass, $h_{r}$ is the radiation damping coefficient, $K = \rho g S$ is the hydrostatic stiffness (with $\rho$ water density, $g$ gravity, and $S$ cross-sectional area), $h_{ex}$ is the excitation force coefficient, and $\eta(t)$ is the ocean wave elevation.

The goal is to design a PTO force profile $\mathcal{F}_{u}(t)$ to maximize energy generation over a finite time horizon. However, the problem is challenging due to two reasons: (1) the ocean wave parameters are unknown to the WEC, and (2) the optimal PTO force profile is inherently periodic and not stationary. We first parametrize this periodic behavior and then relate it to energy generation and ocean wave parameters.

The random ocean waves usually can be represented by linear superposition of a number of harmonic waves \citep{borgman1969ocean}.
To simplify the process of developing an autonomous search strategy of the optimal operation condition of the WEC, it is assumed that the WEC is subject to a wave of a single frequency, i.e. regular wave as referred in the remaining of the paper, in the design stage. The effectiveness and robustness of the developed autonomous control system under irregular waves will be assessed and demonstrated by simulation in Section~\ref{simulation}. That is, the external wave $\eta$ is represented by 
\begin{equation}
  \eta(t)=A \cos(\omega t+B),
  \label{wave}
\end{equation}
where the variable $A$ represents the height of the wave, $\omega$ the frequency 
of $\omega=2\pi/T_{w}$, and with the wave period $T_{w}$, depending on the wave characteristics in different oceans. 
and $B$ the wave phase in the range of from $0$ to $2\pi$. In our setting, all these wave parameter vector $\theta=[A,B,\omega]^T$ are unknown and may change with time. 

%To facilitate our control system design and enforce the periodic behavior of the buoy, it is assumed that an ideal PTO profile is harmonic. This can be justified since the external wave profile is assumed to be harmonic. The low-level PTO implementation is to control the PTO system to follow the desirable harmonic PTO force profile to be specified by the high-level learning and optimization algorithm. Consequently, under a well designed PTO controller, the PTO force $\mathcal{F}_{u}(t)$ can be effectively represented by

Given the periodic nature of both ocean excitation and desirable PTO behavior, we model the ideal PTO force as a harmonic signal:
\begin{equation}
  \mathcal{F}_{u}(t) = A_{u} \cos(\omega_{u} t + B_{u}),
  \label{control force}
\end{equation}
where $\theta_u = [A_u, B_u, \omega_u]^\top$ denotes the control force parameters: amplitude $A_u \in [0, A_{u,\max}]$, phase $B_u \in [0, 2\pi]$, and frequency $\omega_u \in [\omega_{u,\min}, \omega_{u,\max}]$. This representation enables us to describe the non-stationary control action with a compact parameter vector $\theta_u$.

Given the linear nature of system~\eqref{wec motion simple}, the velocity response is the superposition of responses to the wave excitation and PTO force:
\begin{equation}
    v(t) = v_x(t) + v_y(t) = A_x \sin(\omega_u t + B_x) + A_y \sin(\omega t + B_y),
    \label{eqn: velocity formula}
\end{equation}
where $v_x(t)$ and $v_y(t)$ are velocity components induced by PTO and wave forces, respectively. Substituting~\eqref{wave} and~\eqref{control force} into~\eqref{wec motion simple} yields:
\begin{align}
    A_x (m \omega_u - \tfrac{K}{\omega_u}) \cos(\omega_u t + B_x) + A_x h_r \sin(\omega_u t + B_x)
    &= A_u \cos(\omega_u t + B_u), \\
    A_y (m \omega - \tfrac{K}{\omega}) \cos(\omega t + B_y) + A_y h_r \sin(\omega t + B_y)
    &= A h_{ex} \cos(\omega t + B).
\end{align}

Solving for amplitudes and phases, we obtain:
\begin{align}
    A_x &= \frac{A_u}{\sqrt{(m \omega_u - \frac{K}{\omega_u})^2 + h_r^2}}, \\
    B_x &= \arctan\left(\frac{h_r}{m \omega_u - \frac{K}{\omega_u}}\right) + B_u, \\
    A_y &= \frac{A h_{ex}}{\sqrt{(m \omega - \frac{K}{\omega})^2 + h_r^2}}, \\
    B_y &= \arctan\left(\frac{h_r}{m \omega - \frac{K}{\omega}}\right) + B.
\end{align}

\subsection{Average energy generation}
With these expressions, we are now in the position to calculate the energy generated by the WEC with this absorber velocity profile.
The instantaneous power is given by:

\begin{equation}
    P(t) = -\mathcal{F}_u(t) v(t) = P_x(t) + P_y(t),
\end{equation}
where

\begin{equation}
\begin{aligned}
    P_{x}(t)= -\frac{1}{2}A_{u}A_{x}(\sin(2\omega_{u}t+B_{u}+B_{x})-\sin(B_{u}-B_{x}))
    \label{power x}
\end{aligned}
\end{equation}
and
\begin{equation}
\begin{aligned}
    P_{y}(t)=& -\frac{1}{2}A_{u}A_{y}(\sin((\omega_{u}+\omega)t+B_{u}+B_{y})\\
    &-{\rm sin(B_{u}-B_{y}+(\omega_{u}-\omega)t)})
    \label{power y}
\end{aligned}
\end{equation}

Our aim is to optimize WEC's \emph{average} energy generated in a fixed duration $T$ that covers a number of wave periods. We define average power over duration $T$ as:

\begin{equation}
\begin{aligned}
    P_{avg}=\frac{1}{T}\int^{T}_{0}  P_{x}(t)+ P_{y}(t) dt.
\end{aligned}
\label{energy}
\end{equation}

Assuming $T$ spans multiple wave periods, high-frequency sinusoidal terms integrate to zero. The average contributions become:

\begin{equation}
\begin{aligned}
P_{avg,x}(\theta_{u}) = \frac{1}{T}\int^{T}_{0} \frac{1}{2}A_{u}A_{x}\sin(B_{u}-B_{x})dt
     = -\frac{A_{u}^{2}h_{r}}{2((m \omega_{u}-\frac{K}{\omega_{u}})^{2}+h_{r}^{2})}
    \label{energy 1}
\end{aligned} 
\end{equation}
and
\begin{equation}
\begin{aligned}
P_{avg,y}(\theta_{u},\theta)=&\frac{1}{T}\int^{T}_{0} \frac{1}{2}A_{u}A_{y}\sin(B_{u}-B_{y}+(\omega_{u}-\omega)t)dt\\
     =& \frac{A_{u} h_{ex} A}{2T(\omega_{u}-\omega)\sqrt{(m \omega-\frac{K}{\omega})^{2}+h_{r}^{2}}}\\
    & \left(\cos\left(B_{u}-B-{\rm arctan}(\frac{h_{r}}{m \omega-\frac{K}{\omega}})\right) \right.\\
    & \left.-\cos\left(B_{u}-B-{\rm arctan}(\frac{h_{r}}{m \omega-\frac{K}{\omega}})+(\omega_{u}-\omega)T\right)\right)\\
    \label{energy 2}
\end{aligned} 
\end{equation}
where $\theta=[A,B,\omega]$ and $\theta_{u}=[A_{u},B_{u},\omega_{u}]$ denote the amplitude, phase and frequency parameters of the wave and the ideal PTO force trigonometric function \eqref{control force}, respectively. 
%Expanding them into the industrial and physical meaning, 
$P_{avg,x}(\theta_{u})$ in Eq. \eqref{energy 1} is know as the intrinsic mechanical impedance which always be negative \citep{bacelli2020comments}. $P_{avg,y}(\theta_{u},\theta)$ in Eq. \eqref{energy 2} is the kinetic energy caused by reciprocating wave. 

The total average generated energy is obtained by combining Eqs. \eqref{energy 1} and \eqref{energy 2} as 
\begin{equation}
\begin{aligned}
    P_{avg}(\theta_u,\theta)=&-\frac{A_{u}^{2}h_{r}}{2((m \omega_{u}-\frac{K}{\omega_{u}})^{2}+h_{r}^{2})}+ \frac{A_{u} h_{ex} A}{2T(\omega_{u}-\omega)\sqrt{(m \omega-\frac{K}{\omega})^{2}+h_{r}^{2}}}\\
    & \left(\cos\left(B_{u}-B-{\rm arctan}(\frac{h_{r}}{m \omega-\frac{K}{\omega}})\right) \right.\\
    & \left.-\cos\left(B_{u}-B-{\rm arctan}(\frac{h_{r}}{m \omega-\frac{K}{\omega}})+(\omega_{u}-\omega)T\right)\right).
\end{aligned}
\label{Avgenergy}
\end{equation}

By parametrizing the desirable PTO profile with a set of the parameter $\theta_{u}$, as will be shown later, we are able to convert the problem into searching a set of optimal parameters $\theta_u^*$. $h_{ex}$, $h_{r}$, $m$, $K$ in Eq. \eqref{Avgenergy} are always known constants, depending on WEC mechanical and hydrodynamic characteristics. Duration $T$ is a constant to be determined by the designer.
Eq.(\ref{Avgenergy}) establishes how the PTO profile $\theta_u$ and the external wave $\theta$ affect the average generated energy. It plays a vital role not only in deriving the optimal operation profile of the WEC profile as in Section~\ref{sec:optimal}, but more importantly, for auto-optimization control design using the DCEE framework which is described in detail in the Section \ref{DCEE}.

%$\theta=[A,B,\omega]$ representing the key parameters of the wave in unknown ocean environment is estimated based on the measured average energy $P_{avg}$ and the current WEC profile.

\subsection{Optimal periodic operation condition}
\label{sec:optimal}
Based on the average energy generation derived under the harmonic PTO force profile (\ref{Avgenergy}), this subsection is to derive the optimal operation condition and establish the relationship between the optimal PTO operation profile $\theta_u^*= [A_u^*, B_u^*, \omega_u^*]^\top$ and the wave profile $\theta$.  From~\eqref{energy 2}, the maximum of $P_{avg,y}$ is achieved when:
\begin{equation*}
    \sin(B_{u}-B_{y}+(\omega_{u}-\omega)t)\equiv 1, {          \text{     for all     } t}   
\end{equation*}
which is only achieved when 
\   \begin{equation} \label{eq:optimalw}
    \begin{aligned}
   \omega_{u}^{*}=\omega,
    \end{aligned} 
    \end{equation}
and 
\begin{equation}
    B_{u}^*-B_{y}=\frac{\pi}{2}, 
\end{equation}
that is,
\begin{equation} \label{eq:optimalph}
    B_{u}^{*}=B+{\rm arctan}(\frac{h_{r}}{m \omega-\frac{K}{\omega}})+\frac{\pi}{2}.
\end{equation}

Therefore, the optimal PTO operation profile is to match the frequency of the wave. The conclusion is consistent with the results in the frequency-dependent assumption in complex conjugate control \citep{falnes2020ocean} and the MPC approach when regular wave assumption is employed \citep{yetkin2021practical}. It is also consistent with the observation that maximum wave energy power output is achieved when the natural frequency of a WEC matches the frequency of the incoming waves \citep{korde2016hydrodynamic}.

Substituting the derived optimal frequency \eqref{eq:optimalw} and the phase \eqref{eq:optimalph} into~\eqref{Avgenergy} and optimizing w.r.t.\ $A_u$ yields:

    \begin{equation}
    \begin{aligned}
    A_{u}^{*}=\frac{Ah_{ex}\sqrt{(m \omega-\frac{K}{\omega})^{2}+h_{r}^{2}}}{2h_{r}}
    \label{optimal amplitude}
    \end{aligned} 
    \end{equation}

 Therefore, the optimal PTO operational condition $\theta_u^*=[A_u^*,B_u^*,\omega_u^*]^T$ are determined by \eqref{eq:optimalw},\eqref{eq:optimalph} and \eqref{optimal amplitude}. The corresponding optimal float velocity is represented by 
    \begin{equation}
    \begin{aligned}
   v^*&(t)=\frac{Ah_{ex}}{h_{r}}\sin(\omega t+B+2{\rm arctan}(\frac{h_{r}}{m \omega-\frac{K}{\omega}})+\frac{\pi}{2})\\
   &+\frac{Ah_{ex}}{\sqrt{(m \omega-\frac{K}{\omega})^{2}+h_{r}^{2}}}\sin(\omega t+B+{\rm arctan}(\frac{h_{r}}{m \omega-\frac{K}{\omega}})),
    \end{aligned} 
    \end{equation}
 which can be rewritten in a concise form as 
 \begin{equation} \label{eq:velocity}
     v^*(t)=R \sin(\omega t+{\rm arctan}(\frac{h_{r}}{m \omega-\frac{K}{\omega}})+B+\frac{\pi}{2}+\lambda),
 \end{equation}
where $R$ and $\lambda$ are determined by:
\begin{align}
    R &= \frac{A h_{ex}}{\sqrt{(m \omega - \frac{K}{\omega})^2 + h_r^2}} \sqrt{4 + \left(\frac{m \omega - \frac{K}{\omega}}{h_r}\right)^2}, \\
    \lambda &= \arctan\left(\frac{2 h_r}{m \omega - \frac{K}{\omega}}\right).
\end{align}

    It becomes evident that the frequency of WEC aligns with the wave frequency when the optimal operational conditions are achieved. As shown in \eqref{eq:velocity}, the buoy velocity $v$ is out of phase of both the wave and the PTO force under the optimal operation condition, where $\lambda$ is primarily determined by mechanical characteristics $K$ and $m$ and the wave frequency $\omega$.
    %When we disregard the influence of $\lambda$, the phase difference between the buoy velocity $v$ and PTO force $F_{u}$ remains at $90^\circ$. This indicates that the buoy velocity and PTO force are consistently out of phase, aligning with the  energy as described in equation \eqref{energy}. 
    
    The optimal average energy harvest under the optimal operational condition $\theta_u^*$  is given by

    \begin{equation}
    \begin{aligned}
    P_{avg}^*=\frac{A^{2} h_{ex}^{2}}{8 h_{r}},
    \label{optimal energy}
    \end{aligned} 
    \end{equation}
which means that, under the optimal harmonic PTO control profile in the ocean condition of regular wave, the maximum average generated power is determined by wave height $A$ \citep{faltinsen1993sea}.

\section{Autonomous control system for optimal energy generation}
\label{DCEE}

\subsection{Overall architecture of the auto-optimization systems}

\begin{figure}[ht]
\centerline{\includegraphics[width=0.8\textwidth]{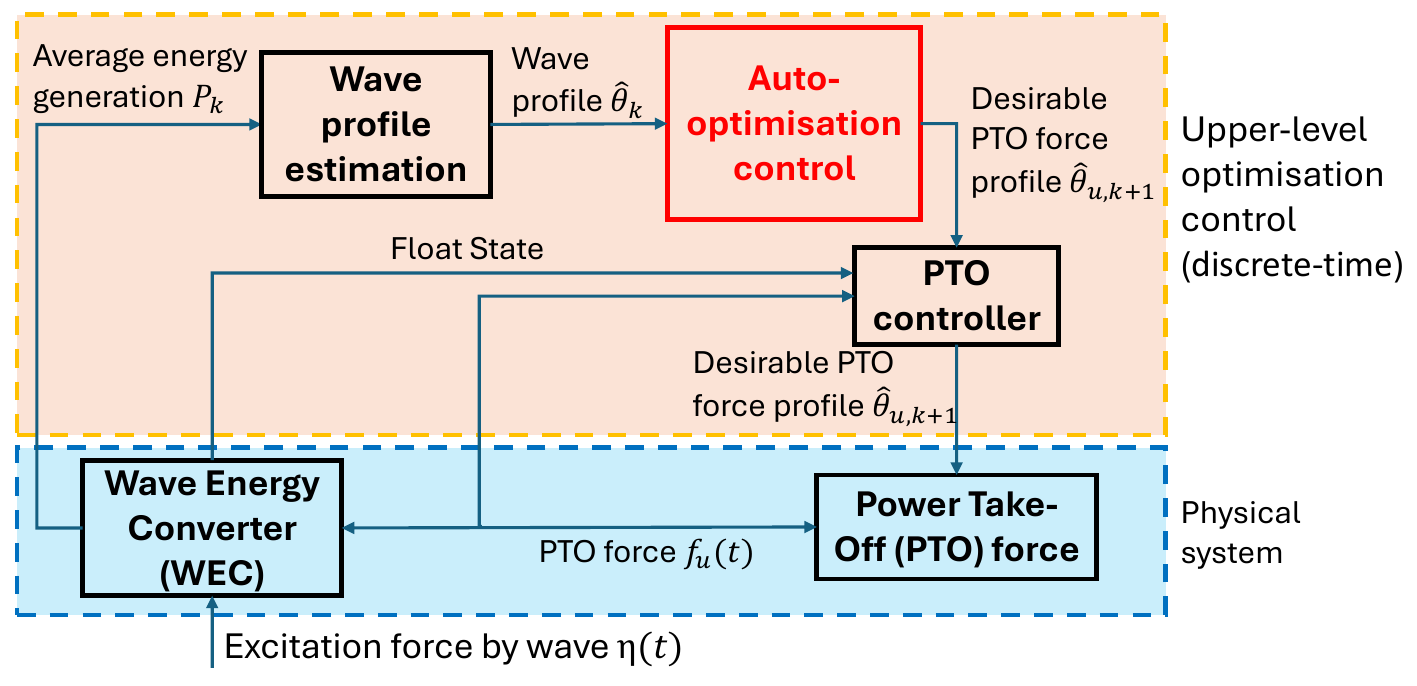}}
\caption{Two-layer control structure}
\label{two level}
\end{figure}

Section~\ref{Problem formulation} provided a parametrization of the WEC's periodic PTO behavior and derived how the ideal profile depends on wave parameters $\theta$. Since wave conditions are unknown and time-varying, the optimal PTO profile must adapt accordingly. To address this, we propose a two-layer adaptive control architecture, illustrated in Figure~\ref{two level}.

High-level controller operates over multiple wave periods. It receives the average output energy $P_{\text{avg}}$, estimates the wave state $\theta$, and computes the next PTO command profile $\theta_u$ using a DCEE algorithm (the following section).
Low-level PTO controller tracks the desired harmonic PTO force $\mathcal{F}_u(t;\theta_u)$, compensating for modeling uncertainties and disturbances to closely follow the high-level command.
The two layers sample at different time scales: the high-level uses averaged energy measurements over a horizon \(T\), while the low-level runs continuously with fast dynamics. This paper focuses on the high-level DCEE design; for low-level implementation details, refer to~\citep{weiss2012optimal}.

\subsection{Dual Control for Exploration and Exploitation (DCEE)} \label{sec:DCEE}

From a general autonomous system perspective, maximization of WEC energy generation in unknown ocean conditions can be considered as to design an auto-optimization control system for an autonomous agent to maximize its reward (i.e. energy) in an unknown environment.  We formulate the problem as to learn a unknown optimal operation condition of WEC through interaction with the unknown ocean condition. Along a similar line, extremum seeking \citep{parrinello2020adaptive} or reinforcement learning \citep{anderlini2020towards} can be applied to the WEC energy generation, and show some promising performance. DCEE as a new autonomous control strategy demonstrates its promising performance due to its active learning capability \citep{chen2022perspective}. It is applied to design the high-level controller in Figure \ref{two level}.

So far DCEE is only applied to systems with stationary optimal operation conditions. 
After parametrizing the periodic behavior of the WEC operation using a set of parameters $\theta_u$, we establish its relationship with the wave condition in Section~\ref{Problem formulation}. With the help of Eq.\eqref{Avgenergy}, we convert the problem of maximizing the energy generation into searching unknown optimal operation condition defined by $\theta_u$. We are now in the position to develop an auto-optimization control system for WEC using the DCEE framework.

When DCEE is applied to WEC, it has two main steps: first estimating the current ocean condition based on all the collected data; and secondly generating the PTO profile that not only drives the WEC to move close to the estimated optimal condition (as calculated in Section~\ref{sec:optimal}) but also actively probes the ocean condition to generate more informative data to support next step's estimation (i.e. active learning). They are presented in the following two subsections.

\subsection{Wave estimation} \label{sec:Bayesian}
The key parameter estimation of the wave condition under regular wave assumption is developed based on the Bayesian estimation. The key feature of Bayesian estimation, different from many other parameter estimation methods maximum likelihood \citep{wyatt1997maximum} and least squares \citep{grainger2021estimating}, is that it provide a probability density function. We are able to not only derive the nominal estimation, but more importantly the uncertainty associated with the nominal estimation. The developed wave parameter estimation method is also vital for the control design part in DCEE where the quantification of the influence of the decision/action on the uncertainty of the future estimation is essential to enable the active learning process. 

At time step $k$, the implemented PTO profile is represented by $\theta_{u,k}=[A_{u,k},B_{u,k},\omega_{u,k}]^T$ and the measurement of average energy is $P_{avg,k}$ at the $k$th step. 

The information state $I_k$ is represented as
\begin{equation}
\begin{aligned}
    I_{k}=[P_{avg,k},\theta_{u,k}]^T
\end{aligned}
\end{equation}
Consequently, the information collected up to time step $k$ is denoted by 
\begin{equation}
\begin{aligned}
\boldsymbol{I_{k}}:=\{I_{1},I_{2},...,I_{k}\}.
\end{aligned}
\end{equation}

The Bayesian estimation is introduced to estimate unknown wave parameter $\theta$. At time $k$, it is denoted as  $\theta_k=[A_k,B_k,\omega_k]^T \in C_{k}$ where $C_{k}$ represents a constraint region. After taking into account the data collected $I(k)$ at $k$, the posterior distribution of the wave parameter estimation $\theta_k$, $p(\theta_{k}|\boldsymbol{I_{k}})$, is updated by Bayesian rule as  
\begin{equation}
\begin{aligned}
    p(\theta_{k}|\boldsymbol{I_{k}})=\frac{p(I_{k}|\theta_{k})p(\theta_{k}|\boldsymbol{I_{k-1}})}{p(I_{k}|\boldsymbol{I_{k-1}})}.
    \label{Bayesian}
\end{aligned}
\end{equation}

More importantly, we also use the same Bayesian rule to evaluate the influence of a candidate PTO profile represented by $\theta_{u,k}$, that is, to predict the posterior distribution of the wave parameter estimation under this candidate PTO profile. When a candidate control input $\theta_{u,k}$ is applied, it predicts the future status of the WEC and the average generated energy using \eqref{Avgenergy}. These hypothetical measurements are fed into \eqref{Bayesian} and the hypothetical posterior distribution of wave parameter estimation, represented as $p(\theta_{k}|\boldsymbol{I_{k+1|k}})$, is calculated. 
%Subsequently, upon generating the new wave parameter estimation $\theta_{k+1|k}$, the average energy measurement is denoted as $\hat{P}_{avg,k+1|k+1}=\mathcal{F}(\theta_{u,k},\theta_{k+1|k})$. 
Consequently, the impact of PTO profile $\theta_{u,k}$ extends beyond merely affecting future energy measurements since it also significantly affects future estimation of the wave condition $\theta$, as will be fully explored in Section~\ref{sec:Activelearning}.

We use the particle filter approach to approximate the Bayesian estimation.
%under the non-linear nature and the uncertainty associated with the estimated results.
In the particle filter implementation, the estimation distribution $p(\theta_k)$ is approximated by a set of weighted random samples $\{\theta_{k}^{i},w_{k}^{i}\}^{N}_{i=1}$ which represents the current state of knowledge about the wave parameters. The posterior distribution $p(\theta_{k}|\boldsymbol{I_{k}})$ is

\begin{equation}
\begin{aligned}
    p(\theta_{k}|\boldsymbol{I_{k}}) \approx \sum^{N}_{i=1} w_{k}^{i} \delta (\Bar{\theta}_{k}-\theta_{k}^{i})
\end{aligned}
\end{equation}
where $\delta(\cdot)$ is a Dirac delta function, $\theta_{k}^{i}$ is a sample representing a potential estimation and $w_{k}^{i}$ is the corresponding normalized weighting such that $\sum_{i=1}^{N} w_{k}^{i}=1$. The nominal estimation of the particle filter is given by 
\begin{equation}
   \Bar{\theta}_{k}=\sum_{i=1}^{N} w_{k}^{i}\theta_{k}^{i}
\end{equation}

The process of recursively calculating the posterior distribution is summarized in Algorithm \ref{alg:pf}.

\begin{algorithm}[ht]
\caption{particle filter of wave parameter estimation}
\label{alg:pf}
\begin{algorithmic}
\Require  average energy measurement $P_{avg,k}$; control action $\theta_{u,k-1}$; prior samples $\{\theta_{k-1}^{i},w_{k-1}^{i}\}^{N}_{i=1}$
    \State Draw samples $\hat{P}_{k-1}^{i}=\mathcal{F}(\theta_{u,k-1},\theta_{k-1}^{i})$
    \State Assign weight $\Bar{w}_{k}^{i}=w_{k-1}^{i} \cdot \frac{p(E_{k}|\theta_{k}^{i})p(\theta_{k}^{i}|\theta_{k-1}^{i})}{q(\theta_{k}^{i}|\theta_{k-1}^{i})}$
    \If{$\theta_{k}^{i} \notin C_{k} $}
        \State $\Bar{w}_{k}^{i}=0$
    \EndIf
    \State Normalize weight $w_{k}^{i}=\Bar{w}_{k}^{i}/\sum_{i=1}^{N} \Bar{w}_{k}^{i}$
    \State Calculate effective sample size $N_{eff}=1/\sum_{i=1}^{N} (w_{k}^{i})^2$
    \If{$N_{eff} < N_{T} $}
        \State Re-sample $\{\theta_{k}^{i},w_{k}^{i}\}^{N}_{i=1}$
    \EndIf
\Ensure posterior samples: $\{\theta_{k}^{i},w_{k}^{i}\}^{N}_{i=1}$
\end{algorithmic}
\end{algorithm}

The likelihood function of the particles that are not in the constraint region $C_{k}$ is set to 0 in an acceptance or rejection process. This step ensures that the estimation results of the particle filter will not have unreasonable deviations due to parameter coupling. More details can be found in Section 5 of \citep{shao2010constrained}.

\subsection{Dual control design and active learning}
\label{sec:Activelearning}

The primary objective of WEC control is to maximize energy harvest. At each step $k$, based on the current estimation of the wave condition $\theta_k$, it can be mathematically formulated as 
\begin{equation}
\begin{aligned}
   \max_{\theta_{u,k}} \mathbb{E}_{\theta_k} \{{P_{avg}} (\theta_{u,k},\theta_{k}) \}
\end{aligned}
\label{eq:maxAverage}
\end{equation}
where the average generated energy $P_{avg}$ is given by \eqref{Avgenergy} and $\theta_k$ is the current estimation of wave parameters given by the Bayesian estimator in Section~\ref{sec:Bayesian}. 
In this context, the PTO force profile denoted by $\theta_{u,k}$ is optimized to achieve the maximum energy generation based on the current estimation of wave parameters $\theta_{k}$. The corresponding set of parameters $\theta_{u,k}$ is used to define the desirable PTO profile that is passed to the low-level PTO controller as its reference.   

It is recognized that the control action affects the interaction between the WEC and the ocean so generate different data that are fed into the estimation process so actually affect the converge rate of the estimation process \citep{chen2022perspective}. Therefore, there is a close interaction between the estimation and the control parts and, by carefully selecting the control action, we are able to generate much informative data to speed up the estimation/learning process which is referred as \emph{active learning} \citep{tong2000active}. DCEE as a practical active learning control method for autonomous systems operating in an unknown environment \citep{chen2022perspective,chen2021dual}. We now apply DCEE to develop WEC control systems with active learning capability. 

To facilitate the development of DCEE \citep{li2023dual} and \citep{li2022concurrent}, we rewrite the optimization problem \eqref{eq:maxAverage} as
\begin{equation}
\begin{aligned}
   \min_{\theta_{u,k}} \mathbb{E}_{\theta(k)} \{(P_{max}-P_{avg,k}(\theta_{u,k},\theta_{k}))^{2} \}
\end{aligned}
\label{eq:miniAverage}
\end{equation}
where $P_{max}$ is a sufficient large number, which can be selected based on the estimated upper bound of WEC average energy generation. Essentially, solving \eqref{eq:miniAverage} and \eqref{eq:maxAverage} give the same optimal average energy.  

Different from the existing DCEE literature, the variables to be optimized are not control inputs, but the key parameters $\theta_u$ used to parametrize the PTO operational profile. Furthermore, the update process of the PTO profile is described by

\begin{equation}
\begin{aligned}
    \theta_{u,k}=\theta_{u,k-1}+\Delta \theta_{u,k}
    \label{control step}
\end{aligned}
\end{equation}
where $\theta_{u,k}=[A_{u,k},B_{u,k},\omega_{u,k}]$ is the PTO profile parameters at current step $k$. The sample duration of the discrete time process is $T$ which is the energy generate duration in \eqref{energy 1}.
$\theta_{u,k}$ is used to define the PTO force profile, which drives the PTO. The control action, $\Delta \theta_{u,k}$, represents the admissible set of the possible changes of the PTO profile. To simplify the search process, three possible actions, namely increase, maintain, and decrease, are considered for each of the three dimensions (amplitude, frequency, and phase) with specified step size for each parameter. For instance, the amplitude change of the PTO force can take a value from $[-\Delta A_{u},0,\Delta A_{u}]$ where $\Delta A_{u}$ is a step size of the amplitude specified by a designer. 
Consequently, there are 27 possible combinations of actions for the three dimensions.

The key idea behind DCEE is that, to generate active learning capability, it is important to predict and assess the influence of the measurements generated from a control action on the future estimation/learning. Therefore, the distinctive feature of the DCEE concept is to not only rely on all the current available data but also the predicted measurement under a candidate control action based on the current understanding of the ocean environment. Hence we augment the current data set $\boldsymbol{I_{k}}$  by a candidate control actin and the responding predicted measurement as
$\boldsymbol{I_{k+1|k}}=[\boldsymbol{I_k},\theta_{u,k|k-1},P_{avg,k+1|k}]$ where $\theta_{u,k|k-1}$ is a candidate PTO profile and $P_{avg,k+1|k}$ is the predicted average energy generation with \eqref{Avgenergy} under the candidate control $\Delta \theta_{u,k}$ and the current estimation of the wave condition $\theta_k$.

Based on the above discussion, we modify the optimal control problem in \eqref{eq:miniAverage} as

\begin{equation}
\begin{aligned}
    &\min_{\Delta \theta_{u,k} \in C_k} \mathcal{J}(\Delta \theta_{u,k})\\
    =&\min_{\Delta \theta_{u,k} \in C_k} \mathbb{E} [(P_{max}-\hat{P}_{avg,k+1|k+1})^{2}|\boldsymbol{I_{k+1|k}}],
\end{aligned}
\end{equation}
subject to \eqref{control step} where $\hat{P}_{avg,k+1|k+1}$ denotes the predicted energy generated under a candidate action $\Delta \theta_{u,k}$. Most importantly, it takes into account both its direct effect of changing the energy generation, but also its indirect effect on the estimation of the wave condition as evident from its calculation process.  The distribution $p(\hat{P}_{avg,k+1|k+1})$ is calculated as follows. For a candidate control action $\Delta \theta_{u,k}$, the average energy $\hat{P}_{avg,k+1|k}(\theta_{u,k|k-1}, \theta_k)$ is predicted using \eqref{Avgenergy} under the corresponding PTO profile $\theta_{u,k|k-1}$ and the current estimated wave parameters $\theta_k$. The control action $\Delta \theta_{u,k}$ (so the PTO profile $\theta_{u,k|k-1}$) and the predicted energy generation $\hat{P}_{avg,k+1|k}(\theta_{u,k|k-1}, \theta_k)$ constitute hypothetical measurements that feed into the parameter estimation algorithm to calculate the predicted posterior of $\theta_{k+1|k}$ as described in Section~\ref{sec:Bayesian}. The distribution of $\hat{P}_{avg,k+1|k+1}$ is calculated by \eqref{Avgenergy} with the PTO profile $\theta_{u,k|k-1}$ and the wave probability distribution $p(\theta_{k+1|k}|\boldsymbol{I_{k+1|k}})$, that is,  
\begin{equation}
    \hat{P}_{avg,k+1|k+1}=P_{avg}(\theta_{u,k|k-1}, \theta_{k+1|k}(\Delta \theta_{u,k})) 
\end{equation} 
where $\theta_{k+1|k}(\Delta \theta_{u,k})$ is used to explicitly indicate the influence of the control action on the future estimation of the wave condition.       

We define $\Bar{P}_{avg,k+1|k+1}$ as the nominal predicted average energy. Let $\Tilde{P}_{avg,k+1|k+1}$ denote the error between each estimate energy to the nominal estimate energy generation, i.e., $\Tilde{P}_{avg,k+1|k+1}=\hat{P}_{avg,k+1|k+1}-\Bar{P}_{avg,k+1|k+1}$. The cost function $\mathcal{J}(\Delta \theta_{u,k})$ can be rewritten as

\begin{equation}
\begin{aligned}
     \mathcal{J}(\Delta \theta_{u,k})=&\mathbb{E}  [(P_{max}-\Bar{P}_{avg,k+1|k+1}-\Tilde{P}_{avg,k+1|k+1})^{2}]\\
    =& \mathbb{E} [(P_{max}-\Bar{P}_{avg,k+1|k+1})^{2}]\\
    &-2\mathbb{E} [\Tilde{P}_{avg,k+1|k+1}(P_{max}-\Bar{P}_{avg,k+1|k+1})^{2}] \\
   & +\mathbb{E}[(\Tilde{P}_{avg,k+1|k+1})^2],
\end{aligned}
\end{equation}
since $\mathbb{E}[\Tilde{P}_{avg,k+1|k+1}]=0$, that

\begin{equation}
\begin{aligned}
    \mathcal{J}(\Delta \theta_{u,k}) =  \mathbb{E} [(P_{max}-\Bar{P}_{avg,k+1|k+1})^{2}] +\mathbb{E}[(\Tilde{P}_{avg,k+1|k+1})^2]
    \label{cost function}
\end{aligned}
\end{equation}

The cost function \eqref{cost function} is composed of two key terms. 
The first term is related to the primary control objective of WEC, which is to guide the WEC towards the estimated optimal operational condition. 
It represents the effort to steer the WEC towards an operation condition that maximizes the expected energy generation (i.e. $\Bar{P}_{avg,k+1|k+1}$). 
The second term of the cost function pertains to the confidence level of the estimation of the optimal WEC profile.
Minimizing this term is essential because it drives the WEC to explore and gather more informative measurements about the environment. 
This, in turn, helps reducing the level of uncertainty associated with the expected maximum energy generation $\Bar{P}_{avg,k+1|k+1}$.  This is why it has an \emph{active learning} capability. To evaluate the cost function and select the best control action from the admissible control set,  the Bayesian estimation algorithm developed in Section~\ref{sec:Bayesian} is re-run with the predicted average energy calculated by \eqref{Avgenergy} under each candidate $\theta_{u,k}$ and the current wave estimation $\theta_{k}$.     

Consequently, the optimization problem \eqref{cost function} for DCEE based WEC control can be formulated as

\begin{equation}
\begin{aligned}
    \min_{\Delta \theta_{u,k} \in C_k} \mathcal{J}(\Delta \theta_{u,k})=&\min_{\Delta \theta_{u,k} \in C_k} [(P_{max}-\Bar{P}_{avg,k+1|k+1})^{2}\\
   & +\sum_{i=1}^{N} (w_{k}^{i}(\hat{P}^{i}_{avg,k+1|k+1}-\Bar{P}_{avg,k+1|k+1}))^2]
   \label{DCEE cost}
\end{aligned}
\end{equation}

To address the local minimum problem in the exploitation part of the cost function caused by non-linear power generation equation \eqref{energy 1}, we introduce the random step size into potential control action design process. Based on the dynamics of the control object \eqref{control step}, the random step size is represented as $ \theta_{u,k}=\theta_{u,k-1}+\alpha_{k} \Delta \theta_{u,k}$, in which $\alpha_{k} \in [\alpha_{min}, \alpha_{max}]$ is a gain which is represented by a random number changing by each step $k$. Upper $\alpha_{min}$ and lower $\alpha_{max}$ bounds are specified by the designer. 

\subsection{Auto-optimization implementation}

The sample duration $T$ is considered to be relatively long, encompassing a good number of wave periods. To improve the accuracy and efficiency of wave parameter estimation, we increase the energy measurement and wave state estimation frequency, thereby obtaining more data from WEC power generation observations, in which the two new average energy durations $T_{1}$, $T_{2}$ are set as ($T_{1}<T_{2}<T$). The wave estimation algorithm is run for the measurements at each duration, and the estimation result from the previous duration is fed into the estimation process of next duration as the prior. In the PTO profile planner in DCEE, the longest duration $T$ is used in the cost function calculation for average energy generation.

\begin{figure}[ht]
\centerline{\includegraphics[width=0.8\textwidth]{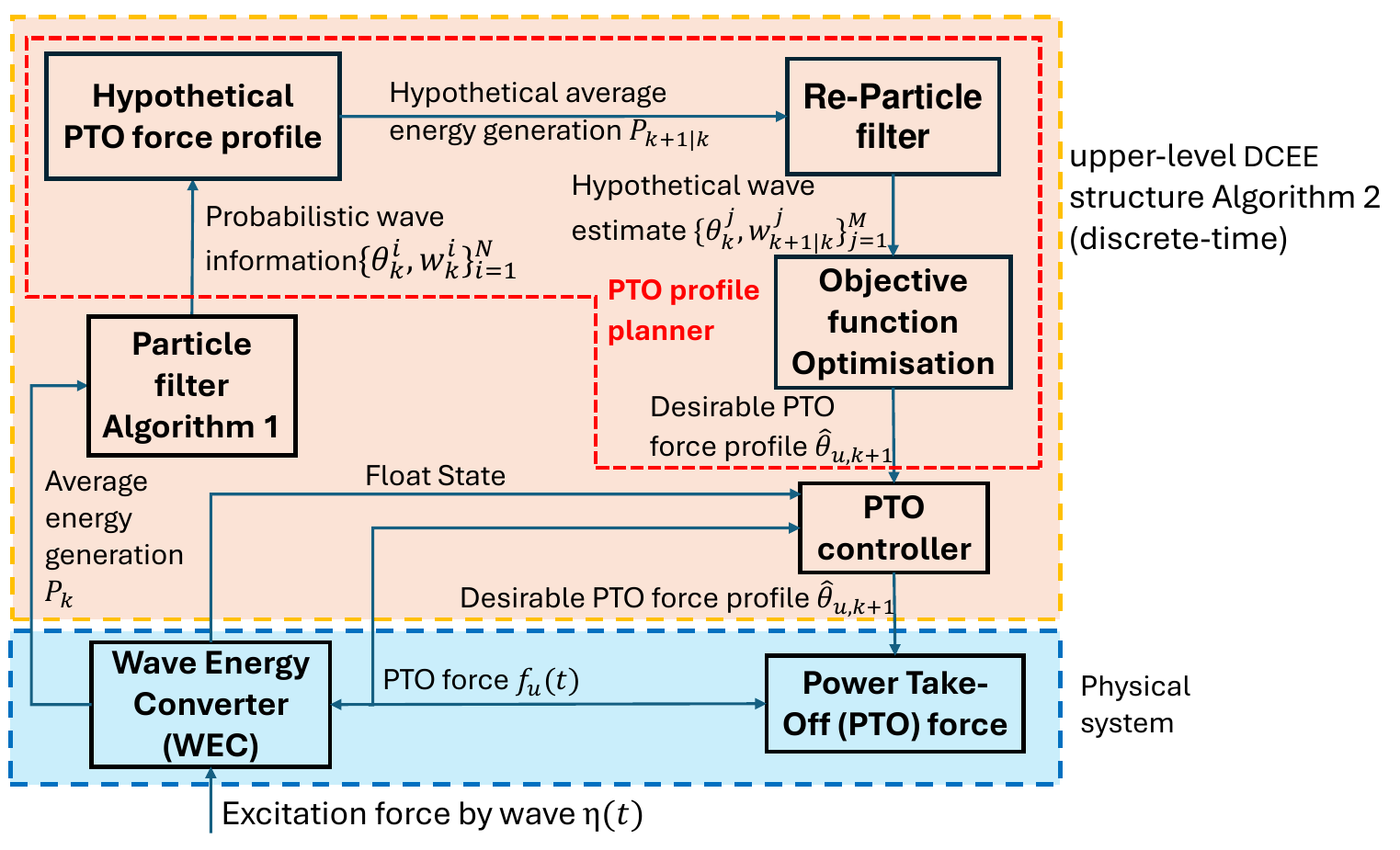}}
\caption{Flow chart of Dual Control of Exploration and Exploitation for wave energy converters}
\label{flow chart}
\end{figure}

Algorithm \ref{alg:wec} encapsulates the implementation of the DCEE for WEC. The architecture and the diagram of auto-optimization of WECs based on DCEE is illustrated in Figure \ref{flow chart}. By observing the average wave energy generation over the duration $T$, the particle filter estimates probabilistic wave parameters. The PTO profile planner provides hypothetical PTO force actions. The predicted power output is obtained using the new control action and updated wave parameter estimation. The hypothetical wave parameter estimation, achieved by reproducing the particle filter process, plays a role in the optimization function. The DCEE algorithm is responsible for acquiring and maintaining the WEC’s optimal operation in an unknown ocean environment. The PTO force profile planner designs an optimal profile for the WEC, specifying the amplitude $A_{u}$, the frequency $\omega_{u}$ and the phase $B_{u}$ of the PTO force. This force is then implemented according to the designed profile, typically using a hydraulic cylinder \citep{weiss2012optimal}. 

\begin{algorithm}[htbp]
\caption{Implementation structure of WEC DCEE}
\label{alg:wec}
\begin{algorithmic}
\Require average energy measurement $P_{avg,1,k}(T_{1})$, $P_{avg,2,k}(T_{2})$, $P_{avg,k}(T)$; control action $\theta_{u,k-1}$; prior samples $\{\theta_{k-1}^{i},w_{k-1}^{i}\}^{N}_{i=1}$ $(k \in [1,K])$
    \State \textbf{Algorithm \ref{alg:pf}}: 
    \State \qquad $\{\theta_{1,k}^{i},w_{1,k}^{i}\}^{N}_{i=1} \gets P_{avg,1,k}, \theta_{u,k-1}, \{\theta_{k-1}^{i},w_{k-1}^{i}\}^{N}_{i=1}$ 
    \State \textbf{Algorithm \ref{alg:pf}}:
    \State \qquad $\{\theta_{2,k}^{i},w_{2,k}^{i}\}^{N}_{i=1} \gets P_{avg,2,k}, \theta_{u,k-1}, \{\theta_{1,k}^{i},w_{1,k}^{i}\}^{N}_{i=1}$ 
    \State \textbf{Algorithm \ref{alg:pf}}:
    \State \qquad $\{\theta_{k}^{i},w_{k}^{i}\}^{N}_{i=1} \gets P_{avg,k}, \theta_{u,k-1}, \{\theta_{2,k}^{i},w_{2,k}^{i}\}^{N}_{i=1}$ 
    \State Down sampling (reduce the computational load)$\{\theta_{k}^{j}\}^{M}_{j=1}$
    \For {each $[\Delta \theta_{u}; \alpha_{k} \Delta \theta_{u}]$}
    \State $\theta_{u,k}=\theta_{u,k-1}+[\Delta \theta_{u}; \alpha_{k} \Delta \theta_{u}]$
    \State Future measurement $P_{avg,k+1|k}^{j}=\mathcal{F}(\theta_{u,k},\theta_{k}^{j},T)$
    \For {j=1,2...M}
    \State Assign weight $\hat{w}_{k+1|k}^{i}=w_{k}^{i} \cdot p(P_{avg,k+1|k}^{j}|\theta_{k}^{i})$
    \State Normalise $w_{k+1|k}^{i}=\hat{w}_{k+1|k}^{i}/\sum_{i=1}^{N} \hat{w}_{k+1|k}^{i}$
    \State Down sampling (for utility calculation)$\{\theta_{k}^{l}\}^{Q}_{l=1}$
    \State Nominal estimation $\Bar{\theta}_{k+1|k}=\sum_{l=1}^{Q} \theta_{k}^{l}$
    \State Measurement $P_{avg,k+1|k+1}^{l}=\mathcal{F}(\theta_{u,k},\theta_{k}^{l},T)$
    \State Nominal $\Bar{P}_{avg,k+1|k+1}=\mathcal{F}(\theta_{u,k},\Bar{\theta}_{k+1|k},T)$
    \State Utility $\mathbb{Y}^{j}=(P_{max}-\Bar{P}_{avg,k+1|k+1})^{2}$
    \State $+\sum_{l=1}^{Q}(\Bar{P}_{avg,k+1|k+1}-P_{avg,k+1|k+1}^{l})^2$
    \EndFor
    \State $\mathbb{E}(\mathbb{Y})=\sum_{j=1}^{M}\mathbb{Y}^{j}$
    \EndFor
   \State $\min \mathcal{J}(\theta_{u,k})=\mathbb{E}(\mathbb{Y})$
\Ensure $\theta_{u,k}$    
\end{algorithmic}
\end{algorithm}

\section{Simulation and evaluation}
\label{simulation}
In this section, the performance of the auto-optimization control scheme based on DCEE is evaluated a WEC system through MATLAB simulation. 
The parameters of the WEC, along with the hydrodynamic coefficients used in the simulation, are listed in Table \ref{tab: WEC parameters}. 
These parameters are adopted from the simulation environment described in \citep{zhang2020model}.

\begin{table}[htbp]
    \centering
    \caption{Parameters of the point absorbers.}
    \begin{tabular}{c c c c}
    \hline
     & parameter & value & unit\\
    \hline
    Stiffness & $K$ & $6.39\times10^{5}$ & $N/m$\\
    Total mass  & $m$ & $8\times10^{3}$ & $kg$\\
    Radiation coefficient & $h_{r}$ & $2\times10^{5}$ & $kg/s$\\
    Excitation coefficient & $h_{ex}$ & $2\times10^{4}$ & $kg/s^{2}$\\
    Control force limit & $\mathcal{F}_{u,max}$ & $21k$ & $N$ \\
    Buoy heave limit & $x_{max}$ & $1$ & $m$\\
    Heave velocity limit & $v_{max}$ & $3$ & $m/s$\\
    \hline
    \end{tabular}
    
    \label{tab: WEC parameters}
\end{table}

The sensor measurement noise is modeled as Gaussian white noise with a mean of zero and a standard deviation of $5\%$ of the signal. The fixed step sizes in \eqref{control step} are set as $\Delta A_{u} = 20 N$, $\Delta B_{u} = 0.002 rad$, and $\Delta \omega_{u} = 0.0005 rad/s$. The sample durations $T_{1}$, $T_{2}$, and $T$ are 20, 30, and 50 seconds, respectively. The number of particles in the particle filter is set to 5,000.

To thoroughly demonstrate the contributions of this paper, the simulation is divided into two parts. The first part focuses on the dynamic performance of the proposed DCEE framework, highlighting its learning capability and convergence toward optimal operating conditions. Extremum Seeking Control (ESC) under the reactive control framework \citep{parrinello2020adaptive} is employed as a benchmark due to its model-free adaptability to uncertain ocean conditions. The second part evaluates steady-state performance in energy harvesting and robustness under irregular waves, with comparison to advanced methods such as Model Predictive Control (MPC) \citep{khedkar2022model} and Bang-Bang control \citep{li2012wave}. Furthermore, we also assess the robustness of our approach under irregular wave conditions.

%provides the operation status when achieves the steady state for showing the advantage of the parametrise formulation.

\subsection{Dynamic performance in learning}

To mimic the real operational environment of WEC systems and demonstrate the self-learning capability of the proposed Bayesian estimation algorithm, a time-varying wave condition profile is simulated as shown below. The ocean wave is initially set as $\cos(0.4\pi t)$, then jumps to $0.7\cos(0.34\pi t+0.5)$ at 10,000 seconds, switches to $0.5\cos(0.32\pi t+0.3)$  after 20,000 seconds, and finally returns to $\cos(0.4\pi t)$. The initial PTO force profile is set to $0$ N.
Extremum Seeking Control (ESC) is introduced as an advanced benchmark control method for comparison with DCEE where the control variables are the tuning of both PTO resistance and reactance \citep{parrinello2020adaptive}.

\begin{figure}[htbp]
\centerline{\includegraphics[width=0.7\textwidth]{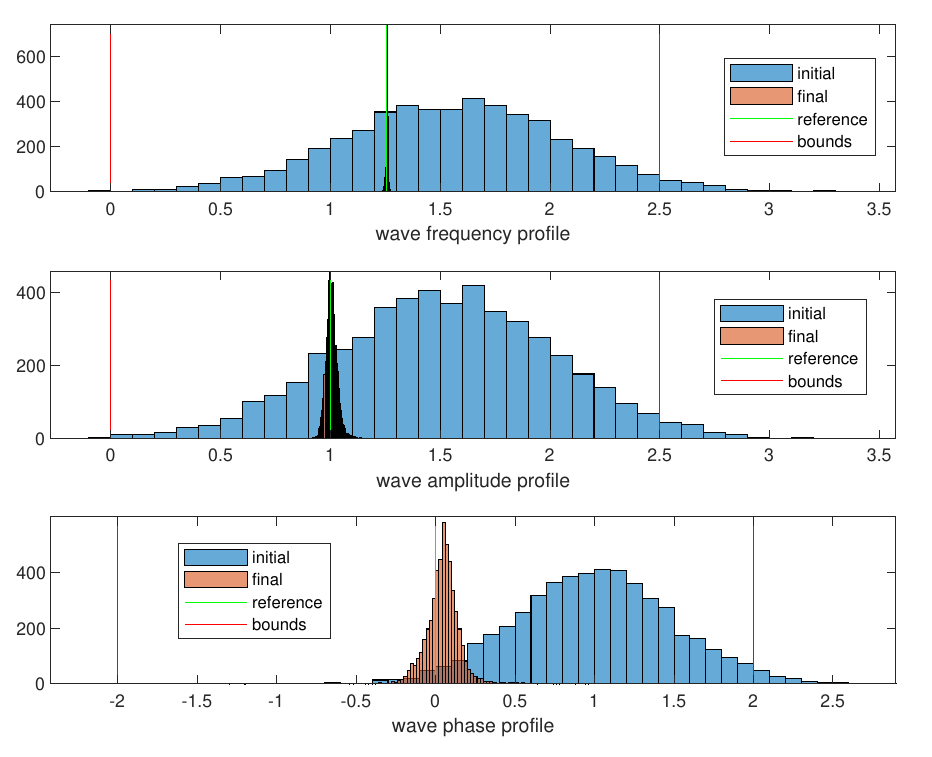}}
\caption{The distribution of particle filter}
\label{distribution}
\end{figure}

\begin{figure}[htbp]
\centerline{\includegraphics[width=0.7\textwidth]{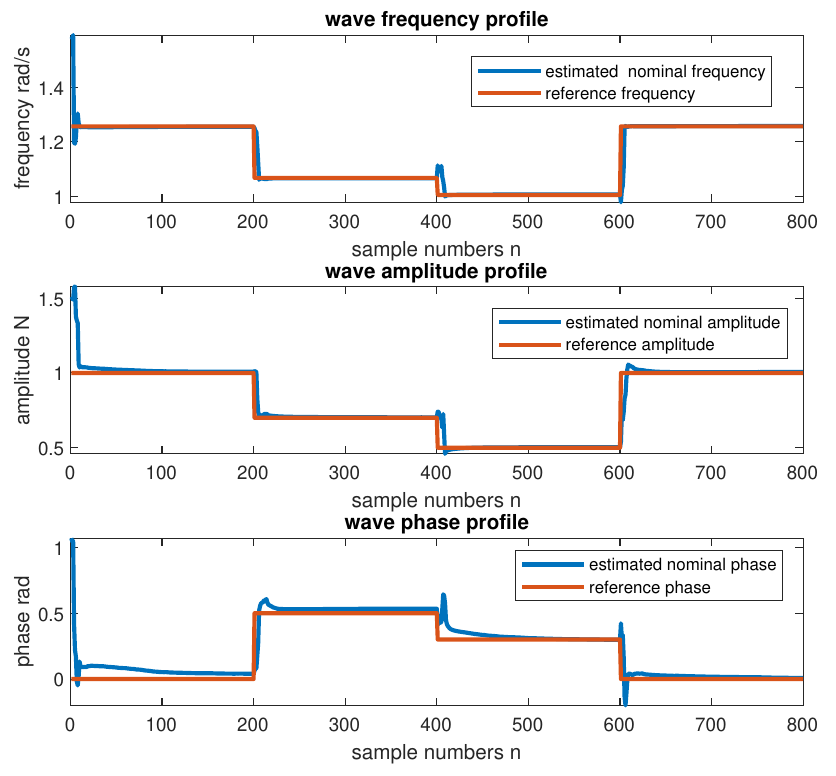}}
\caption{Nominal estimated $\theta_{k}$ using particle filter}
\label{change est}
\end{figure}

\begin{figure}[htbp]
\centerline{\includegraphics[width=0.7\textwidth]{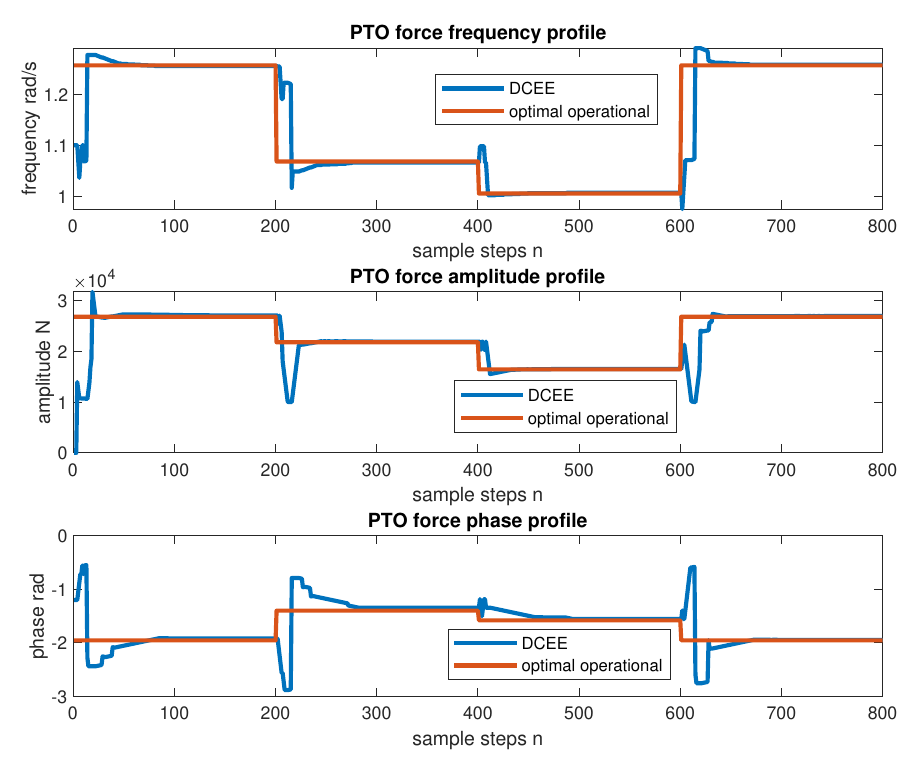}}
\caption{PTO profile $\theta_{u,k}$ using DCEE}
\label{change control}
\end{figure}

\begin{figure}[htbp]
\centerline{\includegraphics[width=0.7\textwidth]{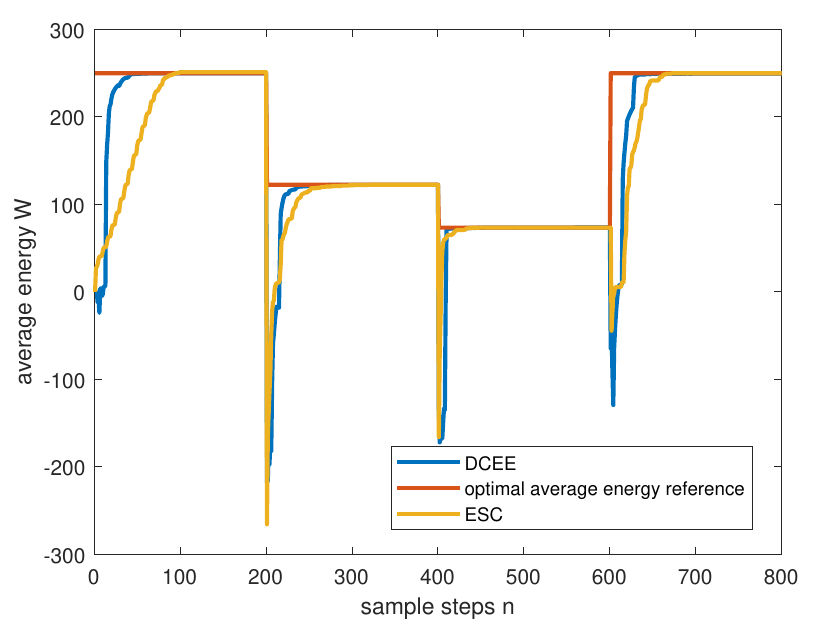}}
\caption{Average energy generation $P_{avg,k}$ using DCEE and ESC}
\label{change energy}
\end{figure}

Figures \ref{distribution}--\ref{change energy} present the simulation results.
The initial setup and final outcomes of the particle within first 10,000 seconds are shown in Figure \ref{distribution}.
The upper and lower bounds of the estimation are set as: $A \in [0,2.5]$, $B \in [-2,2]$ and $\omega \in [0,2.5]$.
Initially, the particles are distributed around $\theta_{0} = [1.5, 1, 1.56]$, following a normal distribution.

Figure \ref{change est} illustrates the evolution of nominal parameter estimates. Despite measurement noise (see Figure \ref{change energy}), the estimation remains robust, demonstrating the effectiveness of the ensemble-based active learning design. The parameter estimates rapidly converge to true values, particularly the wave frequency, which is critical since, according to \eqref{eq:optimalw}, optimal operation requires frequency alignment between the wave and the PTO.

Figure \ref{change control} shows that the PTO profile generated by the DCEE controller adapts swiftly to track optimal conditions. As shown in Figure \ref{change energy}, energy output steadily increases, approaching theoretical maxima. The dual control mechanism, derived from the physical energy objective of the WEC system, inherently balances learning and performance tracking.

Convergence speed is a key metric for evaluating WEC control schemes. A faster convergence enables quicker adaptation and higher energy yield. DCEE exhibits fast convergence by actively balancing exploration (parameter learning) and exploitation (optimal tracking). In contrast, ESC’s convergence speed depends on perturbation signal design and lacks explicit learning. The simulation confirms that DCEE significantly outperforms ESC in both convergence rate and steady-state energy yield.

\subsection{Steady state performance and robustness}

\begin{figure}[htbp]
\centerline{\includegraphics[width=0.7\textwidth]{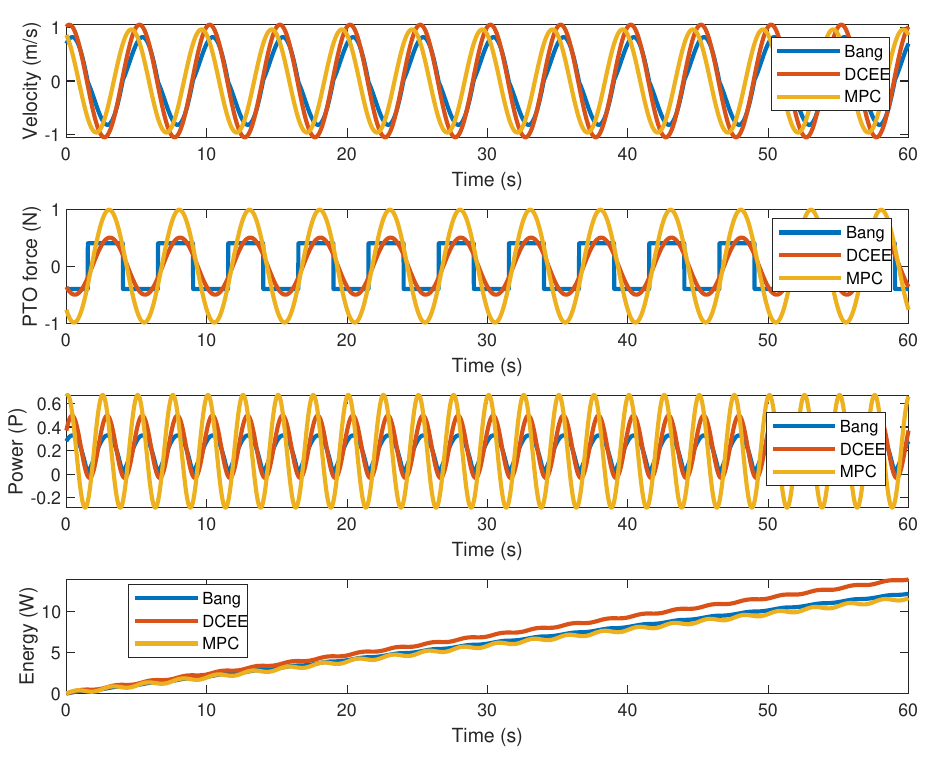}}
\caption{Performance of different WEC control techniques under regular wave}
\label{regular compare}
\end{figure}

\begin{table}[htbp]
    \centering
      \caption{Features of different WEC control techniques under regular wave}
    \begin{tabular}{c c c c}
    \hline
    & 1 & 2 & 3\\
    \hline
    Method & DCEE & MPC & Bang-Bang\\
    \hline
    Energy in $60s$ & $13.95W$ & $11.65W$ & $12.19W$\\
    \hline
    Algorithm complexity & Medium & Complex & Simple\\
    \hline
    Computational consumption & $0.34s$ & $42.68s$ & $\backslash$\\
    \hline
    \end{tabular}
  
    \label{energy efficiency}
\end{table}

In the proposed DCEE-based optimization scheme, the desirable PTO force is parametrized as a sinusoidal signal, as in \eqref{control force}. This section validates this parametrization by comparing the steady-state energy harvesting performance with other control schemes that do not impose such constraints.

Figure \ref{regular compare} compares DCEE, MPC, and Bang-Bang control under regular wave conditions. For fair comparison, it is assumed that DCEE has completed its initial learning phase and has reached its optimal operation. Meanwhile, MPC and Bang-Bang controllers are provided with exact wave and motion information. All signals are normalized for visualization. Results are summarized in Table \ref{energy efficiency}.

The MPC implementation follows \citep{khedkar2022model,richter2012nonlinear}, with real wave height input replacing estimation. The control interval is 0.01 seconds, and the prediction horizon includes 25 steps. Bang-Bang control switches PTO force direction based on the sign of buoy velocity, with no optimization required, making it a practical hardware-friendly solution.

Among the three methods, DCEE achieves the highest energy yield, benefiting from its long-horizon average energy consideration. Although MPC uses ideal wave information in this comparison, its performance is limited by its finite predictive horizon. Extending the horizon improves performance but significantly increases computational cost.

All algorithms were run on the same hardware platform to compare computation demands. While MPC requires frequent updates and extensive computation, DCEE achieves a balance between performance and efficiency. Notably, DCEE's medium complexity results in significantly lower computational cost than MPC, even without considering the cost of wave estimation and forecasting.

\begin{figure}[htbp]
\centerline{\includegraphics[width=0.7\textwidth]{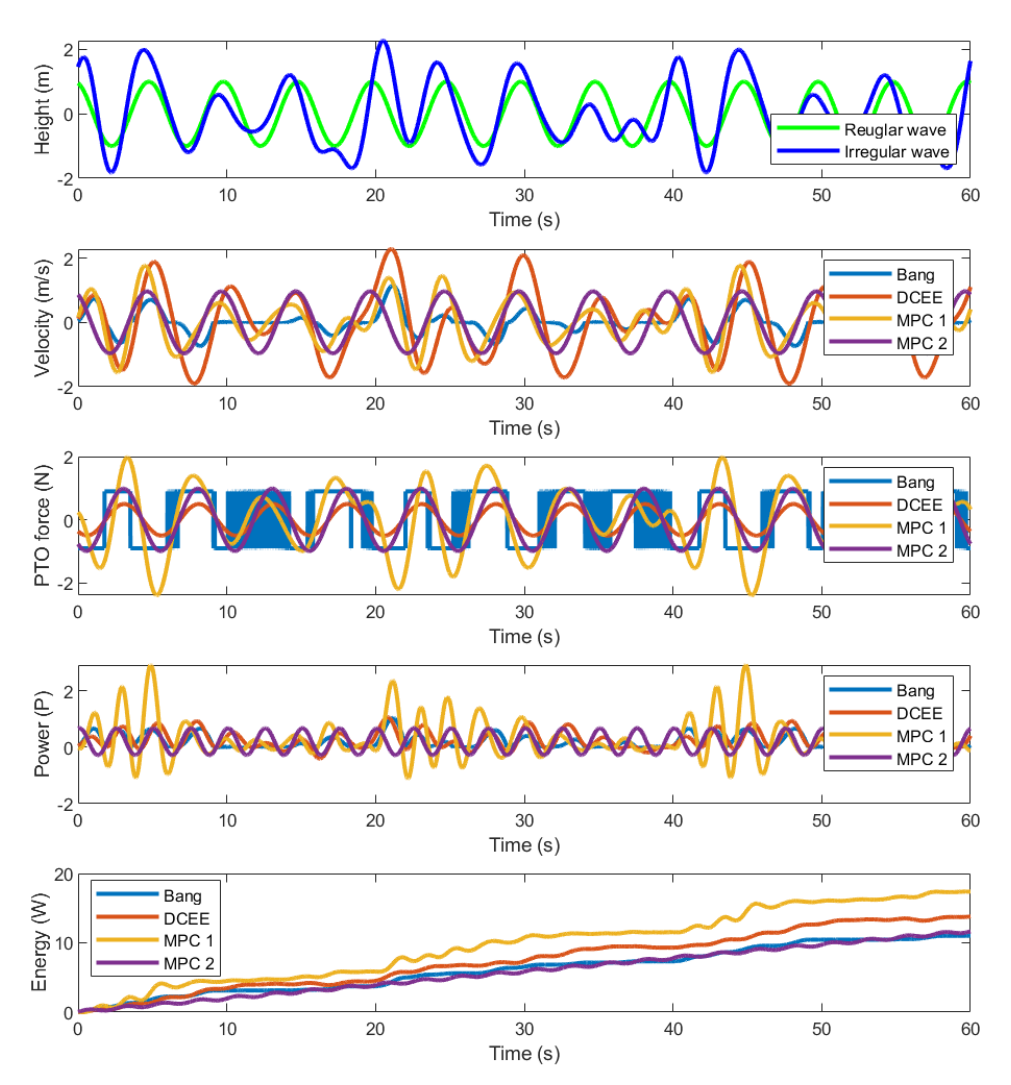}}
\caption{Performance of different WEC control techniques under irregular wave where MPC 1 with exact information of the irregular wave and MPC 2 only with the information of the dominant wave}
\label{irregular compare}
\end{figure}

Figure \ref{irregular compare} evaluates the robustness of the proposed control scheme under irregular wave conditions. The dominant harmonic wave in this case matches that in the regular wave test (Figure \ref{regular compare}). Despite being developed under regular wave assumptions, the DCEE-based controller maintains robust performance under irregular conditions.

Two versions of MPC are used for comparison. MPC 1 assumes access to the full real wave profile, while MPC 2 is limited to the dominant harmonic component. As expected, MPC 1 yields the highest energy, benefiting from complete wave knowledge. DCEE, however, outperforms MPC 2, despite both relying on dominant wave information. The key distinction is that DCEE learns this information autonomously through its estimation and control mechanism, whereas MPC 2 assumes it is pre-provided. Importantly, the strong performance of MPC 1 is achieved at the cost of significantly increased computational burden and an unrealistic assumption of perfect wave information access.

\section{Conclusion}
\label{conclusion}
This paper introduces a promising auto-optimization control framework tailored for WECs to tackle the inherent challenges arising in maximizing WEC energy generation including  non-stationary optimal operation conditions and unknown and changing ocean conditions.
Within this framework, the DCEE approach emerges as a pivotal solution, effectively balancing exploitation and exploration, and promoting active learning.
Through optimizing a cost function consisting of the uncertainty of estimation, the DCEE generates control actions that actively explores the ocean environment in order to reduce uncertainty in the identifying wave parameters and the corresponding optimal PTO operation profile. Simultaneously, it facilitates the real-time tracking of optimal operational conditions of the PTO force.

The simulation results validate the effectiveness and robustness of this novel DCEE based auto-optimization system for WEC, outperforming several existing well-established solutions. DCEE demonstrates excellent learning performance in the presence of time-varying wave profiles. With the same PTO force profile design, DCEE achieves faster convergence compared to ESC algorithms. Under different PTO force profiles, the harmonic-parametrization formulation proposed in this paper generates more energy than other comparable methods. The robustness of the proposed auto-optimization scheme is demonstrated through the tests under irregular wave conditions. Overall, this research represents a significant step toward developing efficient and practical WEC systems in dynamically changing ocean environments, offering a promising approach for wave energy conversion.

 \section*{Acknowledgement}
 This work was supported by the UK Engineering and Physical Sciences Research Council (EPSRC) Established Career Fellowship (EP/T005734/1).

%% If you have bib database file and want bibtex to generate the
%% bibitems, please use
%%
%%  \bibliographystyle{elsarticle-num} 
%%  \bibliography{<your bibdatabase>}

%% else use the following coding to input the bibitems directly in the
%% TeX file.

%% Refer following link for more details about bibliography and citations.
%% https://en.wikibooks.org/wiki/LaTeX/Bibliography_Management
\bibliographystyle{elsarticle-num}
\bibliography{Bibliography}

\begin{thebibliography}{10}
\expandafter\ifx\csname url\endcsname\relax
  \def\url#1{\texttt{#1}}\fi
\expandafter\ifx\csname urlprefix\endcsname\relax\def\urlprefix{URL }\fi
\expandafter\ifx\csname href\endcsname\relax
  \def\href#1#2{#2} \def\path#1{#1}\fi

\bibitem{cruz2007ocean}
J.~Cruz, Ocean wave energy: current status and future prespectives, Springer Science \& Business Media, 2007.

\bibitem{nolan2005optimal}
G.~Nolan, J.~Ringwood, S.~Butler, W.~Leithead, Optimal damping profiles for a heaving buoy wave energy converter, in: ISOPE International Ocean and Polar Engineering Conference, ISOPE, 2005, pp. ISOPE--I.

\bibitem{babarit2006optimal}
A.~Babarit, A.~H. Cl{\'e}ment, Optimal latching control of a wave energy device in regular and irregular waves, Applied Ocean Research 28~(2) (2006) 77--91.

\bibitem{anderlini2018reactive}
E.~Anderlini, D.~Forehand, E.~Bannon, Q.~Xiao, M.~Abusara, Reactive control of a two-body point absorber using reinforcement learning, Ocean Engineering 148 (2018) 650--658.

\bibitem{garcia2017real}
P.~B. Garcia-Rosa, G.~Kulia, J.~V. Ringwood, M.~Molinas, Real-time passive control of wave energy converters using the hilbert-huang transform, IFAC-PapersOnLine 50~(1) (2017) 14705--14710.

\bibitem{parrinello2020adaptive}
L.~Parrinello, P.~Dafnakis, E.~Pasta, G.~Bracco, P.~Naseradinmousavi, G.~Mattiazzo, A.~P.~S. Bhalla, An adaptive and energy-maximizing control optimization of wave energy converters using an extremum-seeking approach, Physics of Fluids 32~(11) (2020) 113307.

\bibitem{anderlini2017reactive}
E.~Anderlini, D.~Forehand, E.~Bannon, M.~Abusara, Reactive control of a wave energy converter using artificial neural networks, International journal of marine energy 19 (2017) 207--220.

\bibitem{richter2012nonlinear}
M.~Richter, M.~E. Magana, O.~Sawodny, T.~K. Brekken, Nonlinear model predictive control of a point absorber wave energy converter, IEEE Transactions on Sustainable Energy 4~(1) (2012) 118--126.

\bibitem{zhan2018adaptive}
S.~Zhan, J.~Na, G.~Li, B.~Wang, Adaptive model predictive control of wave energy converters, IEEE Transactions on Sustainable Energy 11~(1) (2018) 229--238.

\bibitem{faedo2017optimal}
N.~Faedo, S.~Olaya, J.~V. Ringwood, Optimal control, mpc and mpc-like algorithms for wave energy systems: An overview, IFAC Journal of Systems and Control 1 (2017) 37--56.

\bibitem{fusco2011study}
F.~Fusco, J.~V. Ringwood, A study of the prediction requirements in real-time control of wave energy converters, IEEE Transactions on Sustainable Energy 3~(1) (2011) 176--184.

\bibitem{li2014model}
G.~Li, M.~R. Belmont, Model predictive control of sea wave energy converters--part i: A convex approach for the case of a single device, Renewable Energy 69 (2014) 453--463.

\bibitem{ringwood2014control}
J.~V. Ringwood, G.~Bacelli, F.~Fusco, Control, forecasting and optimisation for wave energy conversion, IFAC Proceedings Volumes 47~(3) (2014) 7678--7689.

\bibitem{faedo2021energy}
N.~Faedo, G.~Scarciotti, A.~Astolfi, J.~V. Ringwood, Energy-maximising moment-based constrained optimal control of ocean wave energy farms, IET Renewable Power Generation 15~(14) (2021) 3395--3408.

\bibitem{ringwood2014energy}
J.~V. Ringwood, G.~Bacelli, F.~Fusco, Energy-maximizing control of wave-energy converters: The development of control system technology to optimize their operation, IEEE Control Systems Magazine 34~(5) (2014) 30--55.

\bibitem{wang2024deep}
H.~Wang, V.~Wijaya, T.~Zeng, Y.~Zhang, Deep reinforcement learning-based non-causal control for wave energy conversion, Ocean Engineering 311 (2024) 118860.

\bibitem{chen2021dual}
W.-H. Chen, C.~Rhodes, C.~Liu, Dual control for exploitation and exploration (dcee) in autonomous search, Automatica 133 (2021) 109851.

\bibitem{chen2022perspective}
W.-H. Chen, Perspective view of autonomous control in unknown environment: Dual control for exploitation and exploration vs reinforcement learning, Neurocomputing 497 (2022) 50--63.

\bibitem{hutchinson2019unmanned}
M.~Hutchinson, C.~Liu, P.~Thomas, W.-H. Chen, Unmanned aerial vehicle-based hazardous materials response: Information-theoretic hazardous source search and reconstruction, IEEE Robotics \& Automation Magazine 27~(3) (2019) 108--119.

\bibitem{li2023dual}
Z.~Li, W.-H. Chen, J.~Yang, Y.~Yan, Dual control of exploration and exploitation for self-optimisation control in uncertain environments, arXiv preprint arXiv:2301.11984 (2023).

\bibitem{li2000autonomous}
Z.~Li, W.-H. Chen, J.~Yang, Y.~Yan, Autonomous source seeking and estimation via active information-directed reinforcement learning, Neurocomputing 544 (2023).

\bibitem{bacelli2011control}
G.~Bacelli, J.~V. Ringwood, J.-C. Gilloteaux, A control system for a self-reacting point absorber wave energy converter subject to constraints, IFAC Proceedings Volumes 44~(1) (2011) 11387--11392.

\bibitem{faedo2018finite}
N.~Faedo, Y.~Pe{\~n}a-Sanchez, J.~V. Ringwood, Finite-order hydrodynamic model determination for wave energy applications using moment-matching, Ocean Engineering 163 (2018) 251--263.

\bibitem{fusco2010short}
F.~Fusco, J.~V. Ringwood, Short-term wave forecasting for real-time control of wave energy converters, IEEE Transactions on Sustainable Energy 1~(2) (2010) 99--106.

\bibitem{davis2020wave}
A.~F. Davis, B.~C. Fabien, Wave excitation force estimation of wave energy floats using extended kalman filters, Ocean Engineering 198 (2020) 106970.

\bibitem{yu1995state}
Z.~Yu, J.~Falnes, State-space modelling of a vertical cylinder in heave, Applied Ocean Research 17~(5) (1995) 265--275.

\bibitem{borgman1969ocean}
L.~E. Borgman, Ocean wave simulation for engineering design, Journal of the Waterways and Harbors Division 95~(4) (1969) 557--583.

\bibitem{bacelli2020comments}
G.~Bacelli, R.~G. Coe, Comments on control of wave energy converters, IEEE Transactions on Control Systems Technology 29~(1) (2020) 478--481.

\bibitem{falnes2020ocean}
J.~Falnes, A.~Kurniawan, Ocean waves and oscillating systems: linear interactions including wave-energy extraction, Vol.~8, Cambridge Uuniversity Press, 2020.

\bibitem{yetkin2021practical}
M.~Yetkin, S.~Kalidoss, F.~E. Curtis, L.~V. Snyder, A.~Banerjee, Practical optimal control of a wave-energy converter in regular wave environments, Renewable Energy 171 (2021) 1382--1394.

\bibitem{korde2016hydrodynamic}
U.~A. Korde, J.~Ringwood, Hydrodynamic control of wave energy devices, Cambridge University Press, 2016.

\bibitem{faltinsen1993sea}
O.~Faltinsen, Sea loads on ships and offshore structures, Vol.~1, Cambridge university press, 1993.

\bibitem{weiss2012optimal}
G.~Weiss, G.~Li, M.~Mueller, S.~Townley, M.~Belmont, Optimal control of wave energy converters using deterministic sea wave prediction, Fuelling the Future: Advances in Science and Technologies for Energy Generation, Transmission and Storage 396 (2012).

\bibitem{anderlini2020towards}
E.~Anderlini, S.~Husain, G.~G. Parker, M.~Abusara, G.~Thomas, Towards real-time reinforcement learning control of a wave energy converter, Journal of Marine Science and Engineering 8~(11) (2020) 845.

\bibitem{wyatt1997maximum}
L.~Wyatt, L.~Ledgard, C.~Anderson, Maximum-likelihood estimation of the directional distribution of 0.53-hz ocean waves, Journal of Atmospheric and Oceanic Technology 14~(3) (1997) 591--603.

\bibitem{grainger2021estimating}
J.~P. Grainger, A.~M. Sykulski, P.~Jonathan, K.~Ewans, Estimating the parameters of ocean wave spectra, Ocean Engineering 229 (2021) 108934.

\bibitem{shao2010constrained}
X.~Shao, B.~Huang, J.~M. Lee, Constrained bayesian state estimation--a comparative study and a new particle filter based approach, Journal of Process Control 20~(2) (2010) 143--157.

\bibitem{tong2000active}
S.~Tong, D.~Koller, Active learning for parameter estimation in bayesian networks, Advances in neural information processing systems 13 (2000).

\bibitem{li2022concurrent}
Z.~Li, W.-H. Chen, J.~Yang, Concurrent active learning in autonomous airborne source search: Dual control for exploration and exploitation, IEEE Transactions on Automatic Control 68~(5) (2022) 3123--3130.

\bibitem{zhang2020model}
Y.~Zhang, S.~Zhan, G.~Li, Model predictive control of wave energy converters with prediction error tolerance, IFAC-PapersOnLine 53~(2) (2020) 12289--12294.

\bibitem{khedkar2022model}
K.~Khedkar, A.~P.~S. Bhalla, A model predictive control (mpc)-integrated multiphase immersed boundary (ib) framework for simulating wave energy converters (wecs), Ocean Engineering 260 (2022) 111908.

\bibitem{li2012wave}
G.~Li, G.~Weiss, M.~Mueller, S.~Townley, M.~R. Belmont, Wave energy converter control by wave prediction and dynamic programming, Renewable Energy 48 (2012) 392--403.

\end{thebibliography}

\end{document}